\documentclass[twocolumn, amsmath,amssymb,floatfix,10pt,prd,
superscriptaddress,nofootinbib]{revtex4}
\usepackage[parfill]{parskip}    
\usepackage{graphicx}
\usepackage{amssymb}
\usepackage{epstopdf}
\usepackage{color}
\usepackage{float}
\usepackage{amsmath}
\usepackage{orcidlink}
\usepackage{latexsym}
\usepackage{amsmath}
\usepackage{amssymb}
\usepackage{amsfonts}
\usepackage{blindtext}
\usepackage{subfigure}

\begin{document}

\title{Cosmological FLRW phase transitions and micro-structure under Kaniadakis statistics }

\author{Joaquín Housset}
\email{joaquin.housset.r@mail.pucv.cl}
\affiliation{
Instituto de Física, Pontificia Universidad Católica de Valparaíso, Casilla 4950, Valparaíso, Chile.}

\author{Joel F. Saavedra \orcidlink{0000-0002-1430-3008}}
\email{joel.saavedra@pucv.cl}
\affiliation{
Instituto de Física, Pontificia Universidad Católica de Valparaíso, Casilla 4950, Valparaíso, Chile.}

\author{Francisco Tello-Ortiz
 \orcidlink{0000-0002-7104-5746}}
\email{francisco.tello@pucv.cl}
\affiliation{
Instituto de Física, Pontificia Universidad Católica de Valparaíso, Casilla 4950, Valparaíso, Chile.}

\begin{abstract}

This article studies the thermodynamics phase transitions and critical phenomena of an FLRW cosmological model under truncated Kaniadakis's statistics. The EoS is derived from the corrected Friedmann field equations and the thermodynamics unified first law. To approach thermodynamics, we use an approximation for $K S_{\text{AH}}\ll 1$, on the cosmological equation, and also check the validity of our approximation. Then for different values from Kaniadakis parameter $K$, order $\mathcal{O}(10^{-37})$ for the relic-abundance, $\mathcal{O}(10^{-84})$ for $^{7}$Li-abundance and $\mathcal{O}(10^{-125})$ in the recombination era, the EoS reveals non-trivial critical points where a first-order phase transition occurs a sort of a  van der Waals fluid.
Interestingly, the numerical values of the critical exponents are the same as those of the van der Waals system. Besides, to obtain more insights into the thermodynamics description, the so-called Ruppeiner's geometry is studied through the normalized scalar curvature, disclosing this invariant zone where the system undergoes repulsive/attractive interactions. Near the critical point, this curvature provides again the same critical exponent and universal constant value as for van der Waals fluid.  Despite the similarity, both systems are quite different because the present one considers a relativistic entropy (Kaniadakis entropy) while the Van der Waals gas responds to a classical entropy (Maxwell-Boltzmann).

\end{abstract}

\maketitle

\section{Introduction}

The so-called Kaniadakis's entropy \cite{Kaniadakis:2002zz,Kaniadakis:2005zk} has attracted much interest in recent years. This information measure corresponds to a relativistic generalization of the classical and well-known Boltzmann-Gibbs-Shannon entropy. In \cite{Moradpour:2020dfm}, using the relation between Tsallis's and Kaniadakis's entropies, it was possible to extrapolate this latter into the gravitational context. Specifically, re-written it in terms of the black hole (BH) entropy \cite{Bekenstein:1973ur}. This achievement allowed to explore a series of gravitational phenomena such as observational constraints on holographic dark energy \cite{Luciano:2022knb,Hernandez-Almada:2021aiw,Hernandez-Almada:2021rjs,Drepanou:2021jiv,Nojiri:2022dkr,Nojiri:2022aof,Odintsov:2023vpj,Sania:2023fjx}, generalized second and third laws of thermodynamics \cite{Abreu:2021kwu,Moradpour:2021soz}, slow-roll inflation \cite{Lambiase:2023ryq}, BHs thermodynamics \cite{Luciano:2023bai,Odintsov:2023qfj,Cimidiker:2023kle,Lymperis:2021qty}, micro-canonical and canonical descriptions \cite{Nojiri:2023ikl}, generalized entropy and its microscopic interpretation \cite{Odintsov:2022qnn,Nojiri:2023bom} and general cosmological setups \cite{Lymperis:2021qty,Sheykhi:2023aqa}, to name a few\footnote{For a recent review and  current status of Kaniadakis's statistics including futures perspectives, see for example \cite{Luciano:2022eio}.}.

In this respect, it is worth noticing that most of the mentioned investigations were possible due to the so-called gravity-thermodynamics conjecture \cite{Jacobson:1995ab,Padmanabhan:2003gd,Padmanabhan:2002sha}. This is so because, starting from a purely thermodynamics description, this conjecture allows us to determine the field equations \cite{Lymperis:2021qty,Sheykhi:2023aqa}. Of course, in doing so, some assumptions and prescriptions are necessary. For instance, for spherically symmetric space-times, one needs to consider that the Misner-Sharp (MS) energy \cite{Misner:1964je} obeys the classical thermodynamics relation, that is, it is equal to the product of the density of the matter distribution by the thermodynamics volume. Also, it is necessary to consider that Clausius's relation holds. Etc. 

From the side of static/stationary space-times, for example, BHs, the thermodynamics studies were performed by using the approach given in the seminal articles \cite{Smarr:1972kt,Bardeen:1973gs,Hawking:1975vcx,Gibbons:1976ue} by replacing the Bekenstein's entropy \cite{Bekenstein:1973ur} by the Kaniadakis entropy \cite{Kaniadakis:2002zz,Kaniadakis:2005zk,Moradpour:2020dfm}. On the other hand, for cosmological models, being these space-times  dynamical, the above scheme was done by using 
the thermodynamics unified first law (UFL) \cite{Hayward:1993wb,Hayward:1994bu,Hayward:1997jp}, and of course, changing the event horizon (EH) by the apparent horizon (AH) \cite{Kodama:1979vn,Faraoni:2011hf,Faraoni:2015ula}.

So, taking advantage of these antecedents, 
a natural question arises: Can the usual gravitational entropy (Bekenstein's entropy) be replaced with Kaniadakis's entropy to get non-trivial thermodynamics phenomena for an FLRW cosmological model? In the case of a positive answer, are these phenomena feasible, at least from a purely theoretical point of view? These simple questions arise because in the General Relativity (GR) scenario, it is well-known that the Friedman-Lemaître-Robertson-Walker (FLRW) model does not present thermodynamics phase transitions \cite{Abdusattar:2023pck}. Nevertheless, this situation drastically changes if one introduces further degrees of freedom, such as scalar fields, other than the gravitational ones \cite{Fernandes:2021dsb}. Indeed, any additional field minimally or not minimally coupled to gravity shall modify the field equations and all the thermodynamics-associated quantities, such as the entropy, MS energy, etc. Therefore, one expects (in principle) a more involved description of thermodynamics.

It is clear from the above paragraph that once a modified theory of gravity and its equations are fully provided, and one is able from the Euclidean path integral formulation\footnote{To apply this procedure, first all, one needs to know or obtain a solution of the theory. The path integral is solved on-shell, adding the proper counter-terms.} to obtain the full thermodynamics description. Here, the situation is quite different because the equations of motion assume that the UFL ``always'' holds and takes a specific form of the entropy from the beginning. So, in principle, nothing prevents us from using this fact as the starting point to investigate under what conditions intriguing thermodynamics phenomena are happening.

So, utilizing the above fact, the main aim and motivation of the present research is to realize how Kaniadakis's information measure modifies the thermodynamics behavior of an FLRW cosmological model. To achieve this goal, we use the corrected Friedmann field equations at leading order in the parameter $K$ \cite{Sheykhi:2023aqa}, analytically determining the EoS and its critical points. It should be pointed out that this scheme does not undermine the underlying physics. Moreover, in this way, one can explicitly see how the critical values depend on the $K$ parameter. Although, for the sake of clarity, it is depicted as the full solution too (numerical solution).
The system undergoes a first-order phase transition, similar to the transition of a van der Waals's fluid, however, for large volumes, the ideal ``gas-like'' behavior is more pronounced than the van der Waals case, that is, the pressure decreases more rapidly here. Interestingly, the numerical values of critical exponents coincide with those found for a van der Waals's fluid (in the mean field approximation). 
Moreover, to obtain further insights into the micro-structure of the system, we perform the analysis from the point of view of the so-called Ruppiner's geometry \cite{Ruppeiner:1981znl,Ruppeiner:1983zz,Ruppeiner:1995zz}. This analysis reveals the regions where attractive/repulsive interactions dominate. Near the critical point, the normalized scalar curvature also has the critical exponent value obtained in mean field theory.

The article is organized as follows: Sect. \ref{sec2} briefly reviews Kaniadakis's entropy formulation in the gravitational context, modified Friedmann field equations, the EoS obtained from the UFL, and the corrected Misner-Sharp energy expression in Sect. \ref{sec3} the thermodynamics phase transition, critical exponents, and Ruppeiner's geometry are analyzed in detail, and finally, Sect. \ref{sec4} concludes the present research.

The mostly positive signature $(-,+,+,+)$ is used throughout the article.

\section{Modified Friedmann equations }\label{sec2}

In this section, we review in short how the Friedmann field equations are modified when the well-known Kaniadakis entropy (also known as the $K$-Entropy)\cite{Kaniadakis:2002zz,Kaniadakis:2005zk} is employed. These expressions are obtained through the UFL on an FLRW metric. Moreover, the corresponding expression for the corrected MS energy \cite{Misner:1964je} and the equation of state (EoS) driven thermodynamics description of the FLRW Universe within this modified scenario are given. 

 \subsection{FLRW model and Kaniadakis information measure}

 The original Kaniadakis entropy expression is given by \cite{Kaniadakis:2002zz,Kaniadakis:2005zk}
\begin{equation}\label{kaniadakisentropy}
    S_K = -k_B \sum_{i}n_i\text{ln}_{(K)}n_i \quad \mbox{with} \quad ln_{(K)}x = \frac{x^{K}-x^{-K}}{2K},
\end{equation}
or equivalently \cite{Abreu:2016avj,Abreu:2017fhw,Abreu:2017hiy,Abreu:2018mti}
\begin{equation}
S_K=-k_B \sum_{i=1}^\mathcal{W} \frac{P_i^{1+K}-P_i^{1-K}}{2 K},
\end{equation}
with $P_i$ the probability of a system being in a specific microstate and $\mathcal{W}$ the total configuration number. This information measure is a generalization of the seminal Boltzmann-Gibbs-Shannon entropy. In a more widely context, in the gravitational framework, it has been shown that the Kaniadakis entropy (\ref{kaniadakisentropy}) can be written as follows \cite{Moradpour:2020dfm}
\begin{equation}\label{kani}
    S_K = \frac{1}{K}\text{sinh}{\left(KS_{\text{B-H}}\right)},
\end{equation}
being $S_{\text{B-H}}$ Bekenstein-Hawking (B-H) entropy, in this case the BH entropy given by \cite{Bekenstein:1973ur}
\begin{equation}\label{beke}
    S_{\text{B-H}}=\frac{A}{4}.
\end{equation}
It is worth mentioning that the area $A$ appearing in the above expression corresponds to the area of the EH BHs; however, as we are most interested in studying cosmological models, this area should be replaced by the area of the AH. So, the Eq. (\ref{kani}) shall be expressed as
\begin{equation}\label{kani1}
    S_K = \frac{1}{K}\text{sinh}{\left(KS_{\text{AH}}\right)}, \quad S_{\text{AH}}=\frac{A_{\text{AH}}}{4}, \quad 0<K<1.
\end{equation}
This result is universal in RG; the entropy is always a quarter of the area, independently of the model. This is so because this expression is derived  from the euclidean action using the on-shell lagrangian of the theory, a fundamental object of the theory.

In this case, since there is no associated action principle from which the Kaniadakis's entropy can be derived, one option to find the modified Friedmann field equations is to appeal to the gravitational-thermodynamics conjecture \cite{Jacobson:1995ab,Padmanabhan:2003gd,Padmanabhan:2002sha}. Given a thermodynamic law with corresponding identifications and interpretations of gravitational quantities, the field equations and their corrections can be consistently found. As the FLRW 
\begin{equation}\label{flrw}
    ds^{2}=-dt^{2}+a^{2}(t)\left(\frac{dr^{2}}{1-kr^{2}}+r^{2}d\Omega^{2}\right)
\end{equation}
model is a dynamical system, the associated thermodynamics law to describe it properly, is the so-called UFL \cite{Hayward:1993wb,Hayward:1994bu,Hayward:1997jp}
\begin{equation}\label{UFL}
    \nabla_{i}E=A\psi_{i}+W\nabla_{i}V,
\end{equation}
where the energy--supply is encoded by the vector $A\psi_{i}$ and the density work by $W$. In general, these last two objects are defined by 
\begin{equation}\label{flux}
\psi_{i}\equiv T^{j}_{\,i}\nabla_{j}R+W\nabla_{i}R,
\end{equation}
and 
\begin{equation}\label{workdensity}
    W\equiv-\frac{1}{2}h_{ij}T^{ij},
\end{equation}
respectively. Nevertheless, as the thermodynamics takes place on the AH, to determine this surface, the line element (\ref{flrw}) can be cast as a warped product between a two-dimensional manifold $\mathcal{M}_{2}$ (the $t-r$ plane) and a two-sphere $\mathbb{S}^{2}$
as 
\begin{equation}
    ds^{2}=h_{ij}dx^{i}dx^{j} + R^{2}d\Omega^{2},
\end{equation}
where $h_{ij}$ is the induced metric on the manifold $\mathcal{M}_{2}$ and $R(t,r)\equiv a(t)r$ is the physical radius. So, it is not hard to show that the AH is the solution of the following differential equation \cite{Faraoni:2011hf,Faraoni:2015ula}
\begin{equation}\label{AHsolu}
    h^{ij}\nabla_i R \nabla_j R = 0
\Rightarrow
    R_{\text{AH}}=\frac{1}{\sqrt{H^{2}+\frac{k}{a^{2}}}},
\end{equation}
where $H\equiv \dot{a}/a$ is the Hubble's expansion rate (also known as Hubble's constant). Notice that Latin indexes run over $i,j=t,r$. In this way, the trace of the energy-momentum tensor $h_{ij}T^{ij}$ on the $t-r$ plane orthogonal to the two spheres of symmetry is given by $h_{ij}T^{ij}=p-\rho$, being $p$ the isotropic pressure and $\rho$ the density of the perfect fluid filling the Universe.

Now, projecting the UFL (\ref{UFL}) along the AH, one obtains \citep{Hayward:1997jp} 
\begin{equation}\label{projected}  z^{i}\nabla_{i}E=\frac{\kappa_{\text{HK}}}{8\pi}z^{i}\nabla_{i}A+Wz^{i}\nabla_{i}V,
\end{equation}
where the following identification has been done $A\psi_{i}=\frac{\kappa_{\text{HK}}}{8\pi}z^{i}\nabla_{i}A$  \citep{Hayward:1997jp,Cai:2006rs} and $z^{i}$ is a vector tangent to the AH. This step allows us to identify the so-called  Hayward-Kodama (HK) surface gravity $\kappa_{\text{HK}}$. Taking into account that it is a purely geometric object (independent of the underlying theory), it has the following definition \citep{Hayward:1997jp}
\begin{equation}\label{SG}
  \kappa_{\text{HK}}=\frac{1}{2\sqrt{-h}}\partial_{i}\left(\sqrt{-h}h^{ij}\partial_{j}R\right),  
\end{equation}
where $h\equiv \text{det}(h_{ij})$. Therefore, the temperature $T$ is defined as 
\begin{equation}
    T=\frac{ \kappa_{\text{HK}}}{2\pi}.
\end{equation}
Despite having the same definition as Hawking's temperature \cite{Hawking:1975vcx}, the main difference is that the HK surface gravity is a dynamical object.

After replacing the surface gravity with the temperature in the project UFL (\ref{projected}), the first term on the right-hand side can be recognized as the Clausius relation, $dE=-TdS$, where by using (\ref{kani1}) and some algebraic reduction, one arrives to \cite{Lymperis:2021qty}

\begin{equation}\label{flrw1}
    \text{cosh}{\left[\frac{K\pi}{H^{2}+\frac{k}{a^{2}}}\right]}\left[\dot{H}-\frac{k}{a^{2}}\right]= -4\pi G (\rho + p),
\end{equation}
\begin{equation}\label{flrw2}
    \text{cosh}{\left[\frac{K\pi}{H^{2}+\frac{k}{a^{2}}}\right]}\left[H^{2}+\frac{k}{a^{2}}\right] - K\pi \text{shi}{\left[\frac{K\pi}{H^{2}+\frac{k}{a^{2}}}\right]}=\frac{8\pi G}{3}\rho,
\end{equation}
where $
\operatorname{shi}(x) \equiv \int_0^x \sinh \left(x^{\prime}\right) / x^{\prime} d x^{\prime}
$. 
It should noted that in deriving the above expressions, it was assumed the MS energy  $E=\rho V$, with $V=4\pi R^{3}_{\text{AH}}/3$.  
Obviously, taking the limit $K\rightarrow 0$ the Eqs. (\ref{flrw1})-(\ref{flrw2}) become the original GR Friedmann field equations. Furthermore, these expressions can be formally expressed regarding the AH and its variation through (\ref{AHsolu}).

\subsection{The EoS and the MS energy}

At this stage, we can derive the EoS of the cosmological FLRW model. To achieve this goal, from the Eqs. of motion (\ref{flrw1})-(\ref{flrw2}) the density work (\ref{workdensity}) reads\footnote{From now on we shall fix $G=1$, that is, we are using relativistic geometrized units where also $c=1$.}
\begin{equation}\label{eos}
\begin{split}
    W = \frac{T}{2R_{\text{AH}}}\text{cosh}{(K\pi R^{2}_{\text{AH}})} + \frac{1}{8\pi R^{2}_{\text{AH}}}\text{cosh}{(K\pi R^{2}_{\text{AH}})}&\\ - \frac{3K}{8}\text{shi}{(K\pi R^{2}_{\text{AH}})},
    \end{split}
\end{equation}
where 
\begin{equation}
    T = -\frac{1}{2\pi R_{\text{AH}}}\left(1-\frac{\dot{R}_{\text{AH}}}{2HR_{\text{AH}}}\right)
\end{equation}
has been used. Now, as $W=P$, being $P$ the pressure of the whole system, the Eq. (\ref{eos}) provides 
\begin{equation}\label{eos1}
\begin{split}
    P(v,T) = \frac{T}{v}\text{cosh}{\left(\frac{K\pi v^{2}}{4}\right)} + \frac{1}{2\pi v^{2}}\text{cosh}{\left(\frac{K\pi v^{2}}{4}\right)}&\\ - \frac{3K}{8}\text{shi}{\left(\frac{K\pi v^{2}}{4}\right)},
    \end{split}
\end{equation}
where we have introduced the molar volume $v=2R_{\text{AH}}$, then the $S_{\text{B-H}}$ at the AH becomes $S_{\text{AH}}=\pi v^{2}/4$. As can be seen from the EoS (\ref{eos1}), it is not possible to obtain the information about the critical points $\{v_{c}, T_{c}\}$, where possible phase transitions are happening, in an analytical way. Of course, solving the system numerically will be interesting, including all the information to check under what conditions phase transitions occur. Nevertheless, it is interesting to see how the first corrections influence the thermodynamics description of these modified Friedmann field equations, Eqs. (\ref{flrw1}) and (\ref{flrw2}), what is more, in this way, one can explicitly see how critical values depend on the $K$ parameter (see below). So, expanding the functions $\text{cosh(x)}$ and $\text{shi(x)}$ in the EoS (\ref{eos1}) for $x\equiv K\pi v^{2}/4\ll 1$ and keeping just the leading order, one gets\footnote{The same result can be obtained by expanding the entropy (\ref{kani1}) from the very beginning. See, for example, \cite{Sheykhi:2023aqa} where the Friedmann equations were obtained by using the UFL along with the expanded entropy (\ref{kani1}) at leading order.  }
\begin{equation}\label{eos2}
\begin{split}
P(v,T)=\frac{T}{v}+\frac{1}{2\pi v^{2}}+\left(\frac{\pi^{2}}{32}v^{3}T-\frac{5\pi}{64}v^{2}\right)K^{2}+\mathcal{O}\left(K^{4}\right).
\end{split}
\end{equation}
Next, the criticality conditions for the EoS (\ref{eos2}) are
\begin{equation}\label{criti}
    \left(\frac{\partial P}{\partial v}\right)\bigg{|}_{T}=\left(\frac{\partial^{2} P}{\partial v^{2}}\right)\bigg{|}_{T}=0.
\end{equation}
These conditions lead to
\begin{equation}\label{criticalpoints}
\begin{split}
    v_{c}=\frac{2}{\sqrt{\pi K}}\left(\frac{8-2\sqrt{15}}{\sqrt{15}}\right)^{1/4}, &\\ \quad T_{c}=-\sqrt{\frac{5K}{3\pi}}\left(\frac{15+2\sqrt{15}}{32}\right)^{1/4},
    \end{split}
\end{equation}
and after replacing the critical points (\ref{criticalpoints}) into the EoS (\ref{eos2}), one gets the following value for the critical pressure $P_{c}$,
\begin{equation}\label{cripre}
    P(v_{c},T_{c})=P_{c}=-\frac{\sqrt{75+40\sqrt{15}}}{12\sqrt{2}}K.
\end{equation}

\begin{figure}[H]
    \centering
   \includegraphics[width=0.4\textwidth]{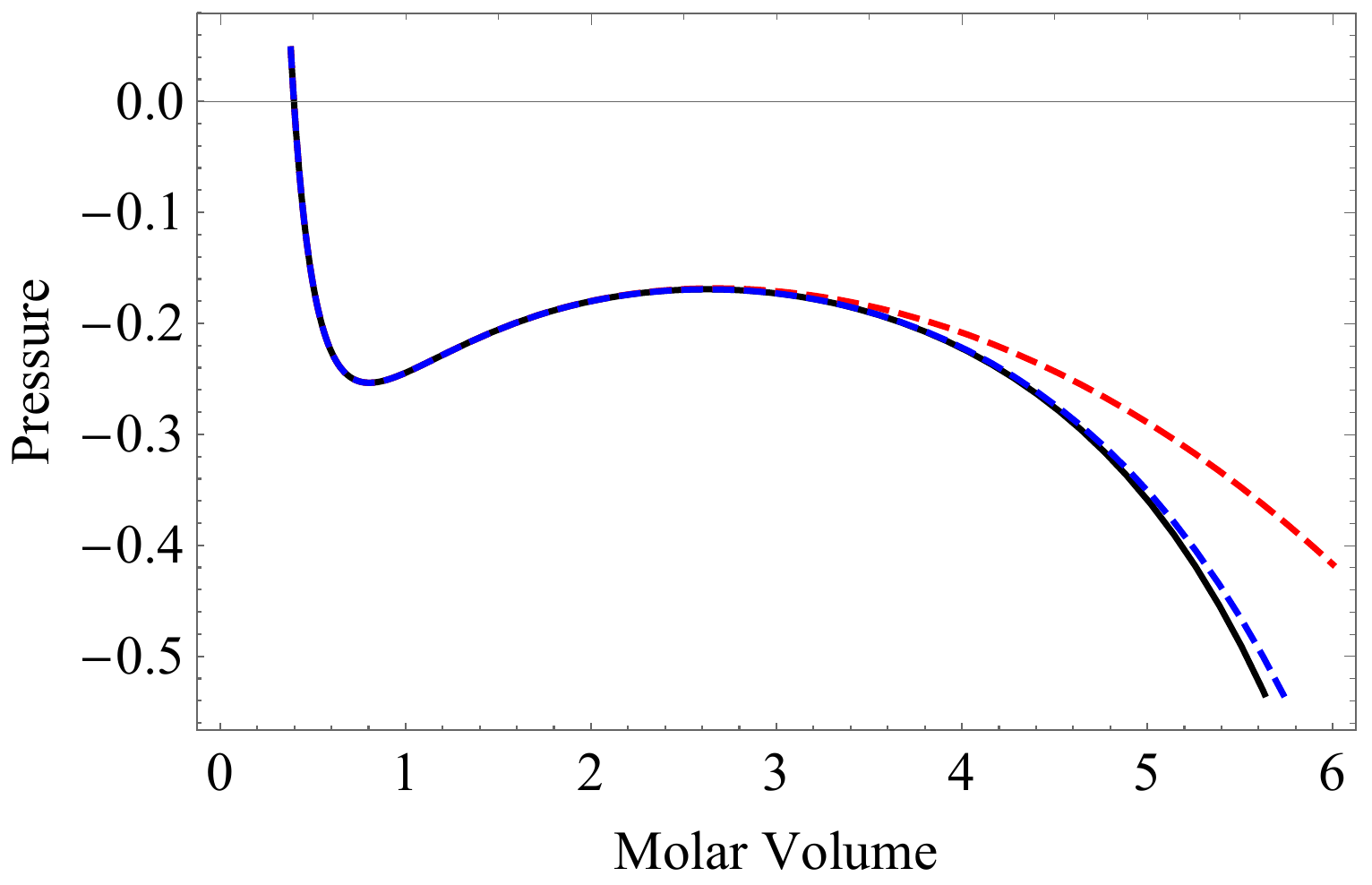}
    \caption{Comparison between the full EoS (\ref{eos1}) and the truncated one (\ref{eos2}), taking into account $K=0.1$ (this value was considered because recent studies \cite{Luciano:2022knb}, suggest that $K$ starts being relevant at order $\mathcal{O}(10^{-1})$). As can be appreciated, both curves differ only at large volumes below the critical temperature. Therefore, the approximated EoS (\ref{eos2}) is valid for the present study. The Solid black line corresponds to the EoS (\ref{eos1}), while the dashed red and blue lines are the expansion at second and fourth orders in $KS_{\text{BH}}$, respectively.}
    \label{figad}
\end{figure}

It is worth mentioning that the system (\ref{criti}) has eight roots. Nevertheless, only one of them (\ref{criticalpoints}) has a physical interpretation because the remaining ones lead to complex values. The Fig. \ref{figad}, shows a comparison between the exact EoS (\ref{eos1}) and the truncated one (\ref{eos2}).\\

Interestingly, the resulting critical temperature $T_{c}$, at which the phase transition occurs, is negative. In this case, the above result can be justified as follows: 
\begin{enumerate}
    \item For those cosmological models presenting an expanding evolution ($\dot{R}_{\text{AH}}>0$), the matter distribution filling the Universe could be, for example (from a purely theoretical point of view) the so-called stiff matter, that is, a matter field satisfying the barotropic EoS $p=\rho$. This matter content was used to study the early Universe stage, leading to a classical bouncing model free from singularities \cite{Oliveira-Neto:2011uhf,Banks:2008ep}. Moreover, the same issue was tackled from the quantum point of view in \cite{Falciano:2007yf}.

    Usually, in this scenario, this kind of matter distribution is accompanied by a negative temperature \cite{Vieira:2016lyj}. Therefore, a cosmological model with a negative temperature is feasible. On the other hand, the final sign of the temperature strongly depends on the sign of the surface gravity and the causal feature of the AH \cite{Binetruy:2014ela,Helou:2015yqa,Helou:2015zma}. In this concern, an expanding cosmology has a past AH leading to $T\propto -\kappa_{\text{HK}}$, while a stiff matter content leads to an outer AH, providing $\kappa_{\text{HK}}>0$, so $T<0$.

    \item Another scenario where a negative temperature drives the cosmological evolution is the case of contracting cosmology ($\dot{R}_{\text{AH}}<0$). In such a case, the matter distribution corresponds to a phantom field \cite{Cruz:2023wtq}, satisfying $p=\omega \rho$ with $\omega<-1$. The main difference with respect to the stiff matter case is that the causal structure of the AH here corresponds to an inner-future horizon. The inner characteristic means $\kappa_{\text{HK}}<0$, whilst the future feature implies 
    $T\propto +\kappa_{\text{HK}}$, thus $T<0$. Nevertheless, this possibility is discarded for the present case since the system (\ref{criti}) has a not real or positive defined volume. 
\end{enumerate}

The previous discussion determines that the present model is dominated by a stiff matter distribution with a negative temperature, corresponding to an expanding cosmological era characterized by an outer-past AH. 

The present of a stiff matter content, is quite relevance, because this situation can be possible (hypothetically speaking), only in the first stages of the Universe's evolution \citep{Banks:2008ep,Chavanis:2014lra}. Therefore, this information allows us to determine what values for the parameter $K$, are plausible to explore thermodynamics phenomena under this hypothetical scenario, following the recent studies \cite{Luciano:2022knb,Hernandez-Almada:2021aiw} about constraints imposed over $K$ due to observational data.

\begin{figure*}
    \centering
    \includegraphics[width=0.32\textwidth]{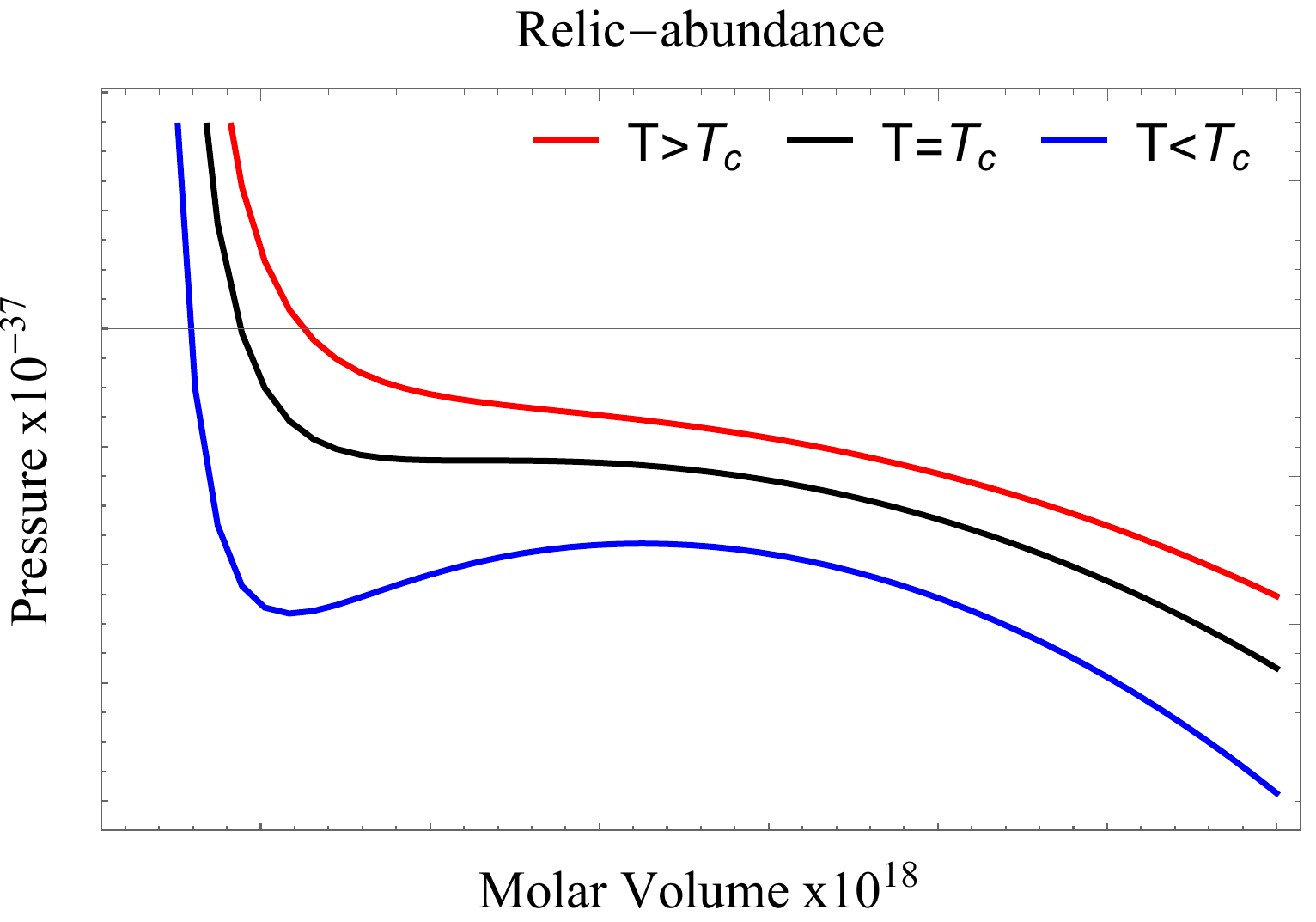}\
    \includegraphics[width=0.32\textwidth]{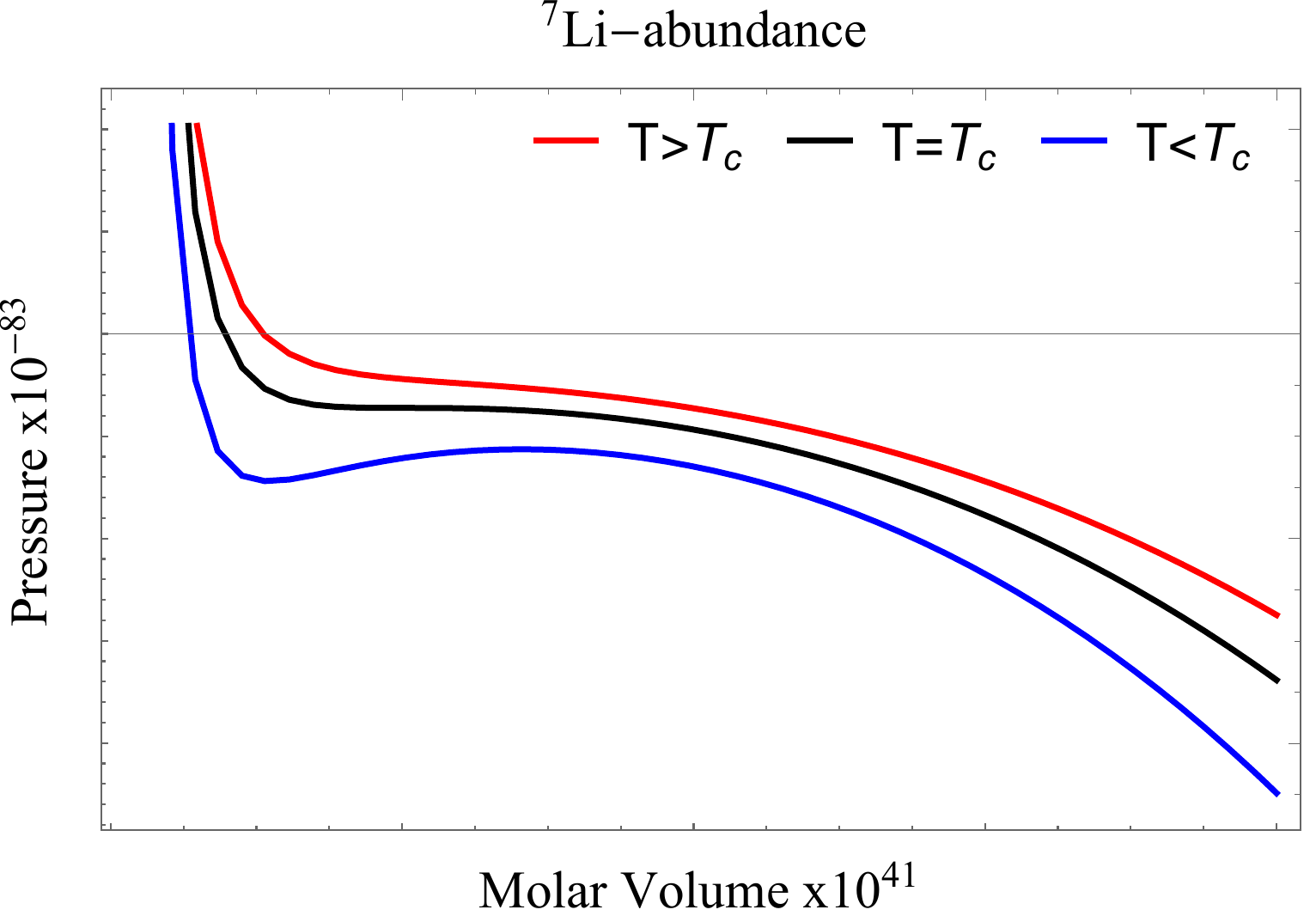}\   
     \includegraphics[width=0.32\textwidth]{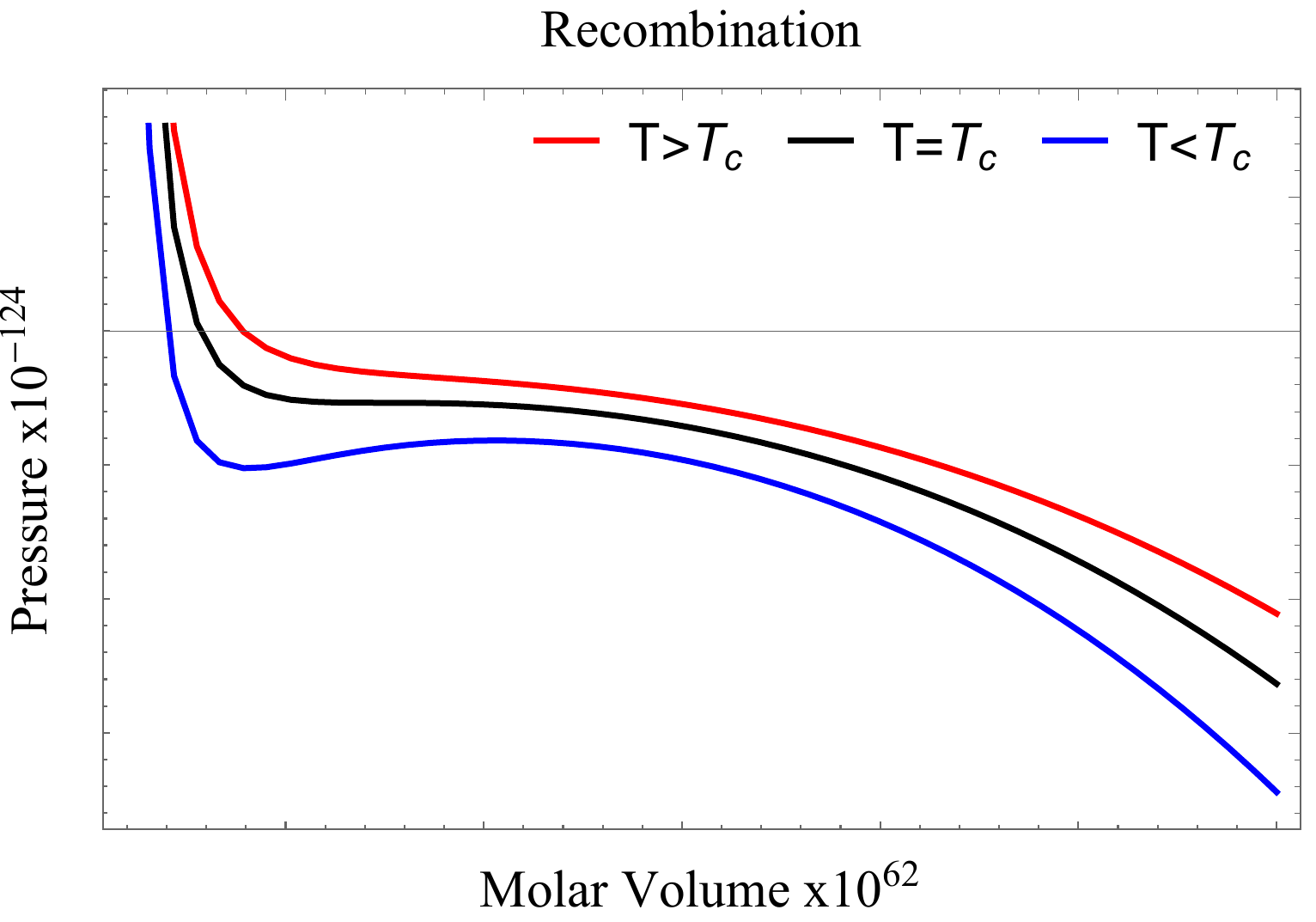}
    \caption{The trend of the EoS (\ref{eos2}) versus the molar volume for different values of the parameter $K$. As reported in \cite{Luciano:2022knb}, the $K$ parameter decreases in magnitude from the early to the late stage of the Universe's history. So, from left to right, the $K$ parameter is of order $\mathcal{O}(10^{-37})$ for the relic-abundance, $\mathcal{O}(10^{-84})$ for $^{7}$Li-abundance and $\mathcal{O}(10^{-125})$ in the recombination era. As can be appreciated, in all cases, the system presents a phase transition similar to what happens in a van der Waals fluid; that is, above the critical temperature, the system behaves as an ideal gas, then below this temperature, there are intermediate states where $\partial P/\partial v>0$, accounting for a phase transition. Despite the similarity, both systems are quite different because the present one takes into account a relativistic entropy while the van der Waals gas responds to a classical entropy. }
    \label{fig1}
\end{figure*}

Finally, we will provide the corrected MS energy to close this section. To achieve this aim, one can proceed in two equivalent ways, that is, i) using $E=\rho V$ or ii) by integrating the UFL (\ref{UFL}). First, this relation is valid since we are dealing with a spherical system, where the volume of the reservoir coincides, in this case, with the physical volume $V$ (the thermodynamics volume). Of course, this assumption was considered in deriving the corrected Friedmann field equations \cite{Lymperis:2021qty,Sheykhi:2023aqa} from the UFL (\ref{UFL}). In the second case, the UFL can be expressed as
\begin{equation}\label{ufl1}
    dE=-TdS+WdV \Rightarrow E=\int(-TdS+WdV).
\end{equation}
The former provides
\begin{equation}
  E= \frac{R_{\text{AH}}}{2 } \text{cosh}{\left(K\pi R^{2}_{\text{AH}}\right)} - \frac{\pi K}{2}R^{3}_{\text{AH}} \text{shi}{\left(K\pi R^{2}_{\text{AH}}\right)}.
\end{equation}
Expanding around $K$, the above expression provides
\begin{equation}\label{MS1}
    E=\frac{R_{\text{AH}}}{2}-\frac{\pi^{2}R^{5}_{\text{AH}}}{4}K^{2}+\mathcal{O}(K^{4}),
\end{equation}
this result also can be obtained from (\ref{ufl1}) inserting the expanded expression of (\ref{kani1}) and performing the algebraic steps. In the limit $K\rightarrow 0$, the expression (\ref{ufl1}) reduces to MS energy corresponding to an entropy $S=A_{\text{AH}}/4$.

\section{Thermodynamics description}\label{sec3}

This section is devoted to analyzing the thermodynamics phase transitions and critical exponents. The phase transition study is performed by employing different perspectives. One using the usual thermodynamics tools \cite{Goldenfeld:1992qy}, that is, by checking the EoS and associated thermodynamics quantities behavior and the second one from the point of view of the so-called Ruppeiner's geometry \cite{Ruppeiner:1981znl,Ruppeiner:1983zz,Ruppeiner:1995zz,Ruppeiner:2013yca,Ruppeiner:2018pgn,Ruppeiner:2023wkq}. This latter shall provide us some insights not only about phase transitions (through the Ruppeiner's scalar curvature), also some information about how the particles of the fluid interact \i.e., an idea about the micro-structure of the system. 
\begin{figure*}
    \centering
    \includegraphics[width=5.8cm,height=5cm,keepaspectratio]{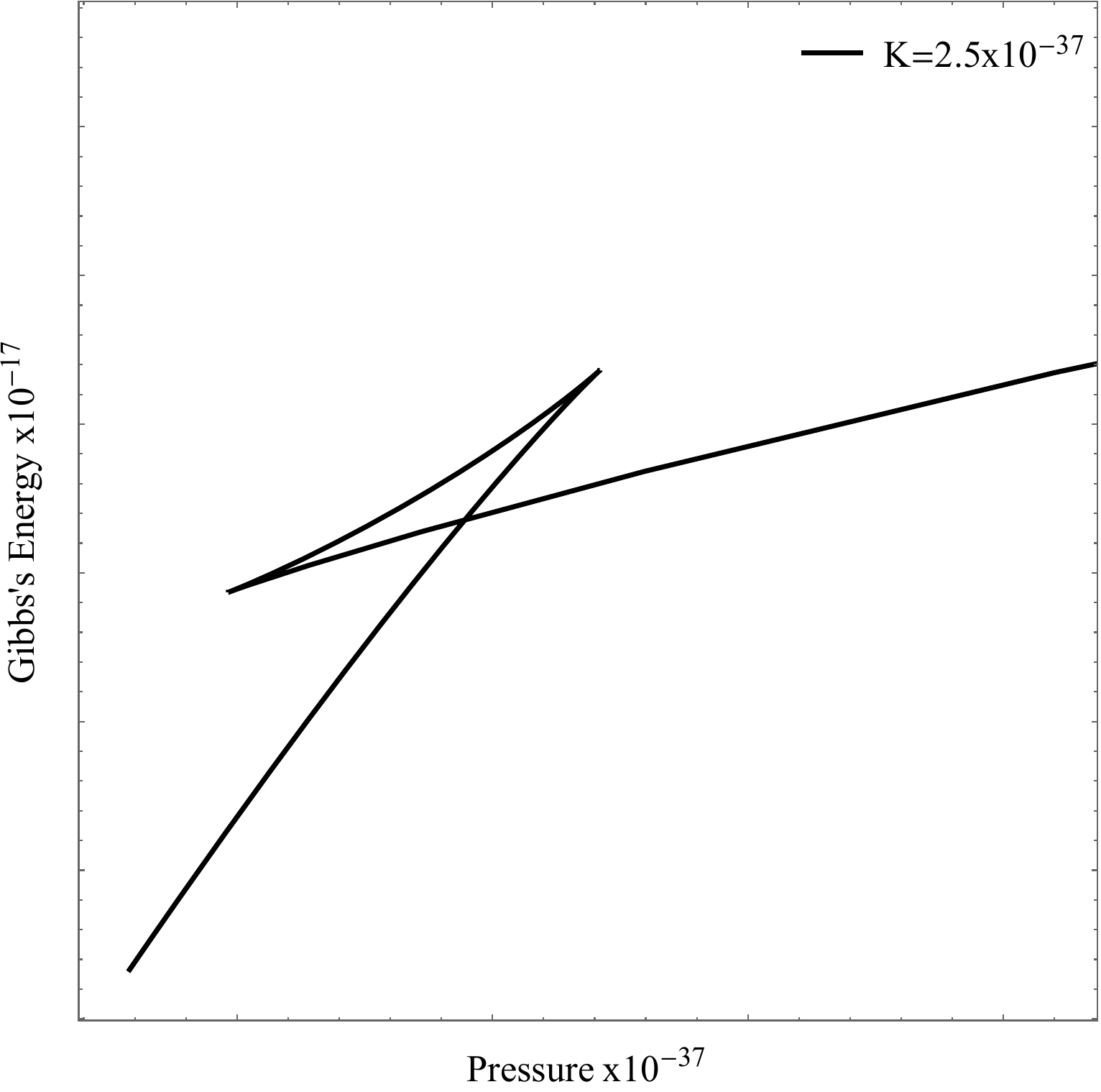}\
    \includegraphics[width=5.8cm,height=5cm,keepaspectratio]{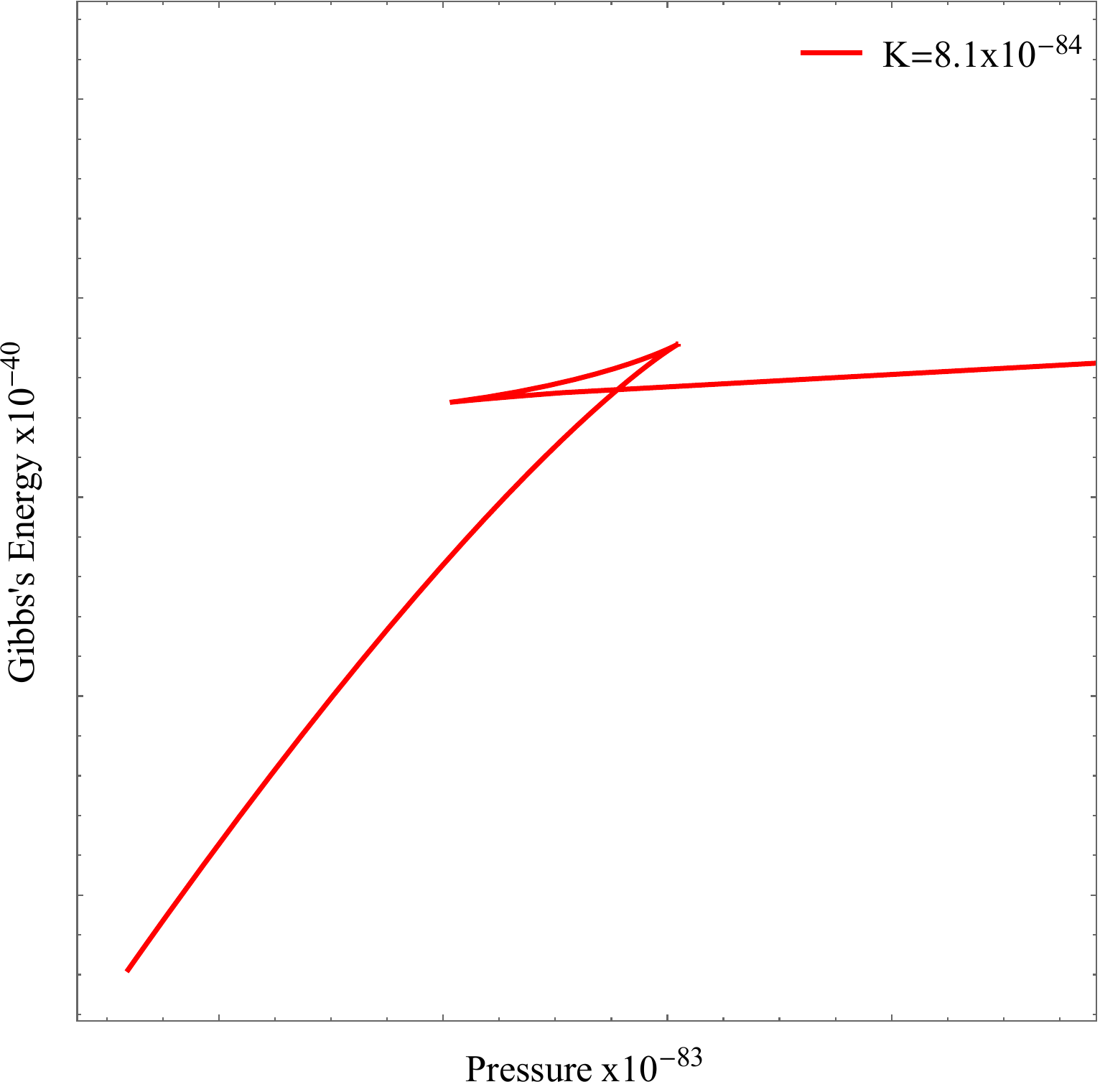}\
    \includegraphics[width=5.8cm,height=5cm,keepaspectratio]{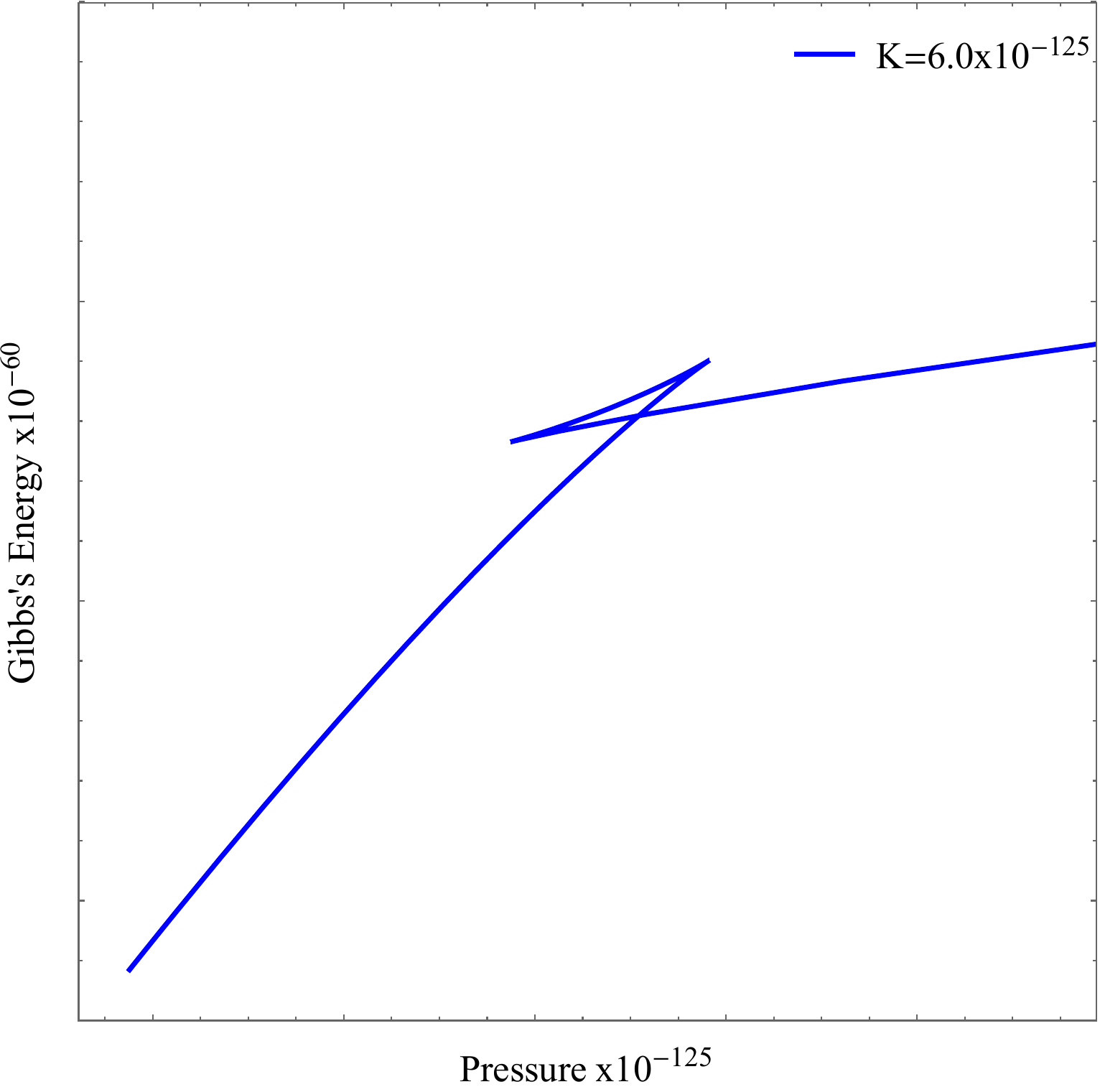}
    \caption{\label{fig2}
    The Gibbs's free energy trend versus the pressure, for $T<T_{c}$. As can be appreciated, this thermodynamics quantity exhibits the classical swallowtail shape, corresponding to a first-order phase transition. From left to right: The relic-abundance, $^{7}$Li-abundance and recombination eras, respectively.}
\end{figure*}

\subsection{Phase transitions}

The EoS (\ref{eos2}) behavior is depicted in Fig. \ref{fig1} (isotherms). Here we have considered those values for the $K$ parameter reported in \cite{Luciano:2022knb}, concerning the early stage of the Universe's evolution. Concretely, we have chosen those values (from left to right) corresponding to relic-abundance, $^{7}$Li-abundance and recombination epochs, where $K$ is of order of $\mathcal{O}(10^{-37})$, $\mathcal{O}(10^{-84})$ and $\mathcal{O}(10^{-125})$, respectively. 
In all cases, the black line shows the critical isotherm, the red one the isotherm above the critical value, and the blue curve depicts an isotherm below the critical temperature, where the system is undergoing a phase transition. The system behaves similarly to a van der Waals fluid in this concern. This is so because the pressure decreases while the volume increases until it reaches a region where $\partial P/\partial v>0$ shows a nonconventional behavior. Those states correspond to thermodynamically unstable states where first-order phase transitions take place. However, one can not say that this system corresponds to a van der Waals fluid because here, the entropy (\ref{kani}) introduces relativistic corrections, while the van der Waals gas is described by a classical entropy, the Boltzmann-Gibbs-Shannon entropy. 

To further corroborate the nature of the phase transition undergone by the system,  
we display in Fig. \ref{fig2} the behavior of the Gibbs's free energy 
\begin{equation}
    \begin{split}
        G(v,T)=\frac{1}{\pi v}+\frac{\pi K^{2}v^{3}}{384}\left(9\pi T v-20\right)-T\text{Log}|v|,
    \end{split}
\end{equation}
versus the pressure (\ref{eos2}), for temperature below the critical one and different values of $K$. The system suffers a first-order phase transition since Gibbs's free energy depicts the standard swallow-tail behavior.
It is important to point out that, as $K$ decreases in magnitude, the swallow-tail also decreases. So, one expects that for the present stage 
$ K \rightarrow 0$, one recovers a non-relativistic FLRW Universe where no phase transitions occur. This behavior can be seen in fig. (\ref{eos2}), where the swallow-tail decreases as $K$ approaches zero, and Gibbs's free energy becomes a decreasing smooth function.  

\subsection{Critical exponents}

Now, we compute the values of the critical exponent associated with this system to get more insights about the system near the critical point. To do so, one needs to analyze the behavior of the heat capacity at constant volume $C_{v}$, the shear viscosity $\eta$, the compressibility $\kappa_{T}$ and pressure $P$ following  
\begin{equation}\label{exponent}
\begin{aligned}
& C_v \sim|\tau|^{-\hat{\alpha}}, \\
& \eta \sim v_l-v_s \sim|\tau|^{\hat{\beta}}, \\
& \kappa_T=-\frac{1}{v}\left(\frac{\partial v}{\partial P}\right)\bigg{|}_T \sim|\tau|^{-\hat{\gamma}}, \\
& P-P_c \sim\left|v-v_c\right|^{\hat{\delta}},
\end{aligned}
\end{equation}
with $\{\hat{\alpha},\hat{\beta},\hat{\gamma},\hat{\delta}\}$ the critical exponents. To obtain the values of these quantities, it is necessary to expand the pressure (\ref{eos2}) around the critical values $\{v_{c}, T_{c}\}$. This expansion produces
\begin{equation}\label{expansion}
\begin{split}
P= P_c+\left[\left(\frac{\partial P}{\partial T}\right)\bigg|_v\right]\bigg|_c\left(T+T_c\right) +\frac{1}{2 !}\left[\left(\frac{\partial^2 P}{\partial T^2}\right)\bigg|_v\right]\bigg|_c &\\ \times\left(T+T_c\right)^2  +\left[\left(\frac{\partial^2 P}{\partial T \partial v}\right)\right]\bigg|_c\left(T+T_c\right)\left(v-v_c\right)&\\ +\frac{1}{3 !}\left[\left(\frac{\partial^3 P}{\partial v^3}\right)\bigg|_T\right]\bigg|_c\left(v-v_c\right)^3+\ldots
\end{split}
\end{equation}
Notice that the criticality conditions (\ref{criti}) have been used in the above expansion. Also, those terms corresponding to the expansion around the critical temperature are proportional to $(T+T_{c})$ and not to $(T-T_{c})$ since our thermodynamics system has a negative absolute temperature. Now, introducing the dimensionless variables $\tau\equiv T/T_{c}+1$ and $\omega\equiv v/v_{c}-1$ and the reduced variables
\begin{equation}\label{reducequantities}
    \tilde{P}\equiv \frac{P}{|P_{c}|}, \quad \tilde{v}\equiv \frac{v}{v_{c}}, \quad \tilde{T}\equiv \frac{T}{T_{c}},
\end{equation}
leading to 
\begin{equation}\label{reduceseos}
\begin{split}
    \tilde{P}=\frac{1}{2\sqrt{255-60\sqrt{15}}\tilde{v}^{2}}\bigg(3\sqrt{3}-6\sqrt{5}\tilde{v}\tilde{T}+3\bigg(5\sqrt{3}&\\-4\sqrt{5}\bigg)\tilde{v}^{4}+\left(6\sqrt{5}-8\sqrt{3}\right)\tilde{v}^{5}\tilde{T}\bigg),
    \end{split}
\end{equation} 
the expression (\ref{expansion}) reads
\begin{equation}\label{expansion1}
\tilde{P}=1+A \tau+B \tau \omega+C\tau\omega^{2}+D \omega^3 +\ldots,
\end{equation}
where the coefficients $\{A,B,C,D\}$ are given by
\begin{equation}
\begin{aligned}
A & =-\frac{4}{35}(5+2 \sqrt{15}), & B & =-\frac{4}{35}(15- \sqrt{15}) \\
C & =2\sqrt{\frac{3}{5}}, & D & =\frac{2}{7}(1-\sqrt{15}).
\end{aligned}
\end{equation}
Notice that the reduced EoS (\ref{reduceseos}), is independent of the parameter $K$. Besides, as the reduce variables are absolute variables and the pressure $P$ is not positive defined everywhere, then to keep the coherence with the non-reduced (\ref{eos2}).

As the reduced pressure (\ref{reduceseos}) is linear in $\tilde{T}$, the heat capacity at constant volume $C_{v}$ is zero, implying $\hat{\alpha}=0$. Now, employing Maxwell's area rule (to determine the remaining critical exponents)
\begin{equation}
    \tilde{P}^{*}\left(\omega_{s}-\omega_{l}\right)=\int^{\omega_{s}}_{\omega_{l}}\tilde{P}d\omega,
\end{equation}
from (\ref{expansion1}) one obtains
\begin{equation}\label{intmax}
\begin{split}
 \tilde{P}^{*}\left(\omega_{s}-\omega_{l}\right)=\int^{\omega_{l}}_{\omega_{s}}\left(1+A \tau+B \tau \omega+C\tau\omega^{2}+D \omega^3\right)d\omega.
\end{split}
\end{equation}
In the above integral expression, $\omega_{s}$ and $\omega_{l}$ stand for the so-called small and large volumes in two different phases. In between these two,
we have a mixture of both of them. Next, the vapor's endpoint and the liquid's starting point have the same pressure. Similarly, here the pressure does not change, that is, $\tilde{P}^{*}=\tilde{P}_{l}=\tilde{P}_{s}$. Therefore, from (\ref{expansion1}) one gets
\begin{equation}\label{equalpre}
    B\tau\left(\omega_{l}-\omega_{s}\right)+C\tau\left(\omega^{2}_{l}-\omega^{2}_{s}\right)+D\left(\omega^{3}_{l}-\omega^{3}_{s}\right)=0.
\end{equation}
On the other hand, the integral in the right member of (\ref{intmax}) provides
\begin{equation}\label{intmax1}
\begin{split}
   \left(1+A\tau\right)\left(\omega_{l}-\omega_{s}\right)+ \frac{B\tau}{2}\left(\omega^{2}_{l}-\omega^{2}_{s}\right)+\frac{C\tau}{3}\left(\omega^{3}_{l}-\omega^{3}_{s}\right)&\\+\frac{3D}{4}\left(\omega^{4}_{l}-\omega^{4}_{s}\right)=\tilde{P}(\omega_{s},\tau)\left(\omega_{s}-\omega_{l}\right).
    \end{split}
\end{equation}
Solving the set of Eqs. (\ref{equalpre})-(\ref{intmax1}) for $\omega_{s}$ and $\omega_{l}$, one obtains the following result
\begin{equation}\label{smalllarge}
\begin{aligned}
\tilde{v}_{s} &=1+\sqrt{\hat{A}_{s}\tau+\hat{B}_{s}\tau^{2}}+\hat{C}_{s}\tau, \\
\tilde{v}_{l} &=1+\sqrt{\hat{A}_{l}\tau+\hat{B}_{l}\tau^{2}}+\hat{C}_{l}\tau,
\end{aligned}
\end{equation}

\begin{figure*}
    \centering
    \includegraphics[width=0.32\textwidth]{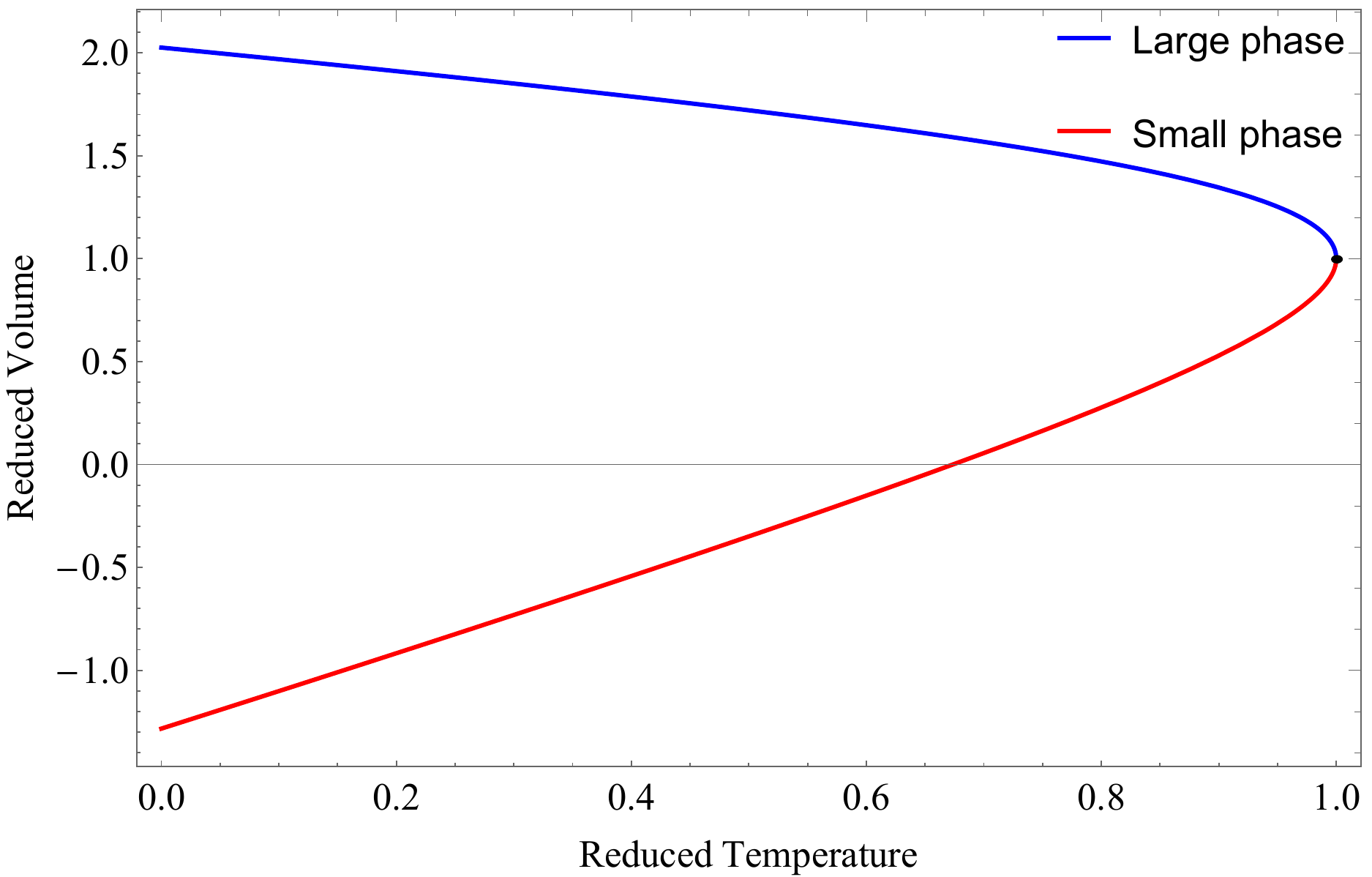}\
    \includegraphics[width=0.32\textwidth]{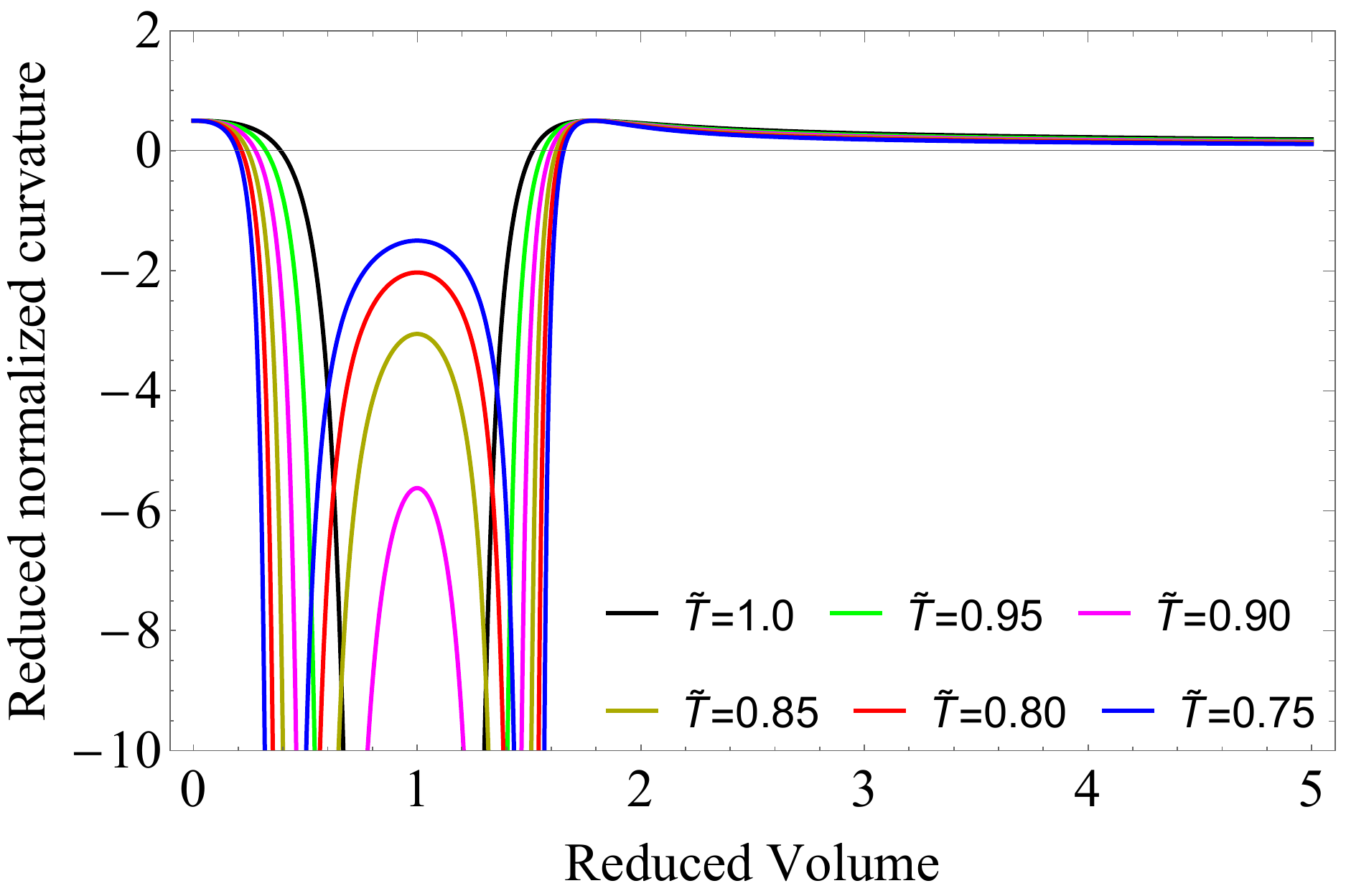}
    \
    \includegraphics[width=0.32\textwidth]{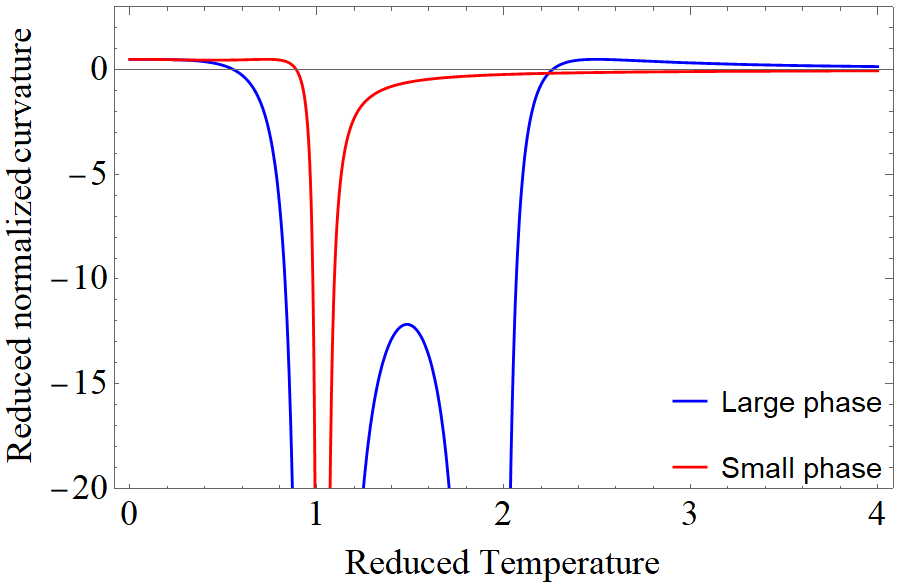}
    \caption{ \textbf{Left panel}: The coexistence locus, plotted regarding the reduced volume and temperature. The blue line corresponds to the so-called large phase, while the red one is to the so-called small phase. The black dot is the critical point where these two phases meet. \textbf{Middle panel}: The trend of the reduced normalized curvature versus the reduced volume for different values of $\tilde{T}$ above the critical temperature. As seen (from the right to the left), it is positive, presenting the system as a repulsive interaction. After some point, it becomes negative, meaning the presence of an attractive interaction. Finally, it retakes positive values. This fact accounts for the existence of phase transition. \textbf{Right panel}:  Again, the behavior of the reduced normalized curvature, but this time as a function of the reduced temperature at the small and large phases.  }
    \label{fig3}
\end{figure*}
where $\{\hat{A}_{i},\hat{B}_{i},\hat{C}_{i}\}$, $i=s,l$, are numerical coefficients. So, from the second relation of Eq. (\ref{exponent}) and taking the lower order in $\tau$ of expressions (\ref{smalllarge}), one gets
\begin{equation}
    \eta\sim v_{l}-v_{s}\sim (\omega_{l}-\omega_{s})v_{c}\sim |\tau|^{1/2}.
\end{equation}
Therefore, the critical exponent $\hat{\beta}$ equals $1/2$. For the critical exponent corresponding to the compressibility $\kappa_{T}$, using (\ref{expansion1}) up first order, 
\begin{equation}
\left(\frac{\partial \tilde{P}}{\partial \tilde{v}}\right)\bigg{|}_{\tau} \simeq \frac{B}{{v}_c} \tau,
\end{equation}
where $\partial\omega/\partial\tilde{v}=1/v_{c}$ has been used. Then, the third expression in (\ref{exponent}) becomes 
\begin{equation}
\kappa_T \simeq \frac{1}{B \tau} \sim \tau^{-1},
\end{equation}
yielding to $\hat{\gamma}=1$. Lastly, by evaluating (\ref{expansion1}) at the critical temperature, this gives us
\begin{equation}
    \tilde{P}-1 \sim \omega^3 \sim\left(\tilde{v}-1\right)^3,
\end{equation}
providing $\hat{\delta}=3$. Collecting all the critical exponent values one has: $\{\hat{\alpha},\hat{\beta},\hat{\gamma},\hat{\delta}\}=\{0,1/2,1,3\}$. These values are the same numerical values predicted for a van der Waals's gas. In some sense, this system can be viewed as an inverted van der Waals's system, that is, the stable branches of a van der Waals's fluid are permuted by unstable branches for a Kaniadakis's fluid\footnote{Of course, we are dealing with Kaniadakis's fluid considering only the first non-trivial and relevant corrections.} and vice-versa. Exhibiting both systems a first-order phase transition with some similarities.

Finally, it is essential to highlight the behavior of the reduced volume against the reduced temperature at the coexistence locus. As can be appreciated in Fig. \ref{fig3} (left panel), the coexistence curve exhibits the so-called small (red line) and large (blue line) phases, with the black dot as the critical point at which these two phases meet. As the phase transition takes place for those values $\tilde{T}<T_{c}$, the coexistence curves cover this region. 

\subsection{Ruppeiner's Geometry}

Ruppeiner's geometry or geometrothermodynamics \cite{Ruppeiner:1981znl,Ruppeiner:1983zz,Ruppeiner:1995zz}, is a powerful tool to understand the micro-structure of a thermodynamical system better. To do so, it is necessary to compute the scalar curvature (Ricci's scalar) associated with the following thermodynamics line element 
\begin{equation}\label{ruppe}   
\begin{split}
dl^{2}=g_{\mu\nu}dx^{\mu}dx^{\nu}\Rightarrow d l^2=\frac{1}{T}\bigg(-\frac{C_v}{T} d T^2&\\+\left(\frac{\partial P}{\partial v}\right)\bigg|_T d v^2\bigg).
\end{split}
\end{equation}
To obtain the above line element expressed in terms of thermodynamics variables, one needs to endow the geometry with thermodynamical meaning as follows: i) the space-time coordinates $x^{\mu}$ should be functions of the internal energy $U$ and the molar volume $v$, ii) the metric tensor $g_{\mu\nu}$ should be related with the entropy $S$ $g_{\mu\nu}=-\frac{\partial^{2} S}{\partial x^{\mu}\partial x^{\nu}}$ and iii) express the first thermodynamic law in terms of the entropy $S$ and then use thermodynamics relation among the variables.

Using the reduced pressure (\ref{reduceseos}), the associated reduced Riccis's scalar $\tilde{\mathcal{R}}$ to the $(\tilde{v},\tilde{T})$-plane described by the line element (\ref{ruppe}) is given by
\begin{equation}\label{curvature}
    \tilde{\mathcal{R}}(\tilde{v},\tilde{T})=\frac{\left(\sqrt{3}-\tilde{A}\tilde{v}^{4}\right)\left(\sqrt{3}-\tilde{A}\tilde{v}^{4}-2\sqrt{5}\tilde{v}\tilde{T}+\tilde{B}\tilde{v}^{5}\tilde{T}\right)}{2C_{v}\left(\sqrt{3}-\tilde{A}\tilde{v}^{4}-\sqrt{5}\tilde{v}\tilde{T}+\tilde{B}\tilde{v}^{5}\tilde{T}\right)^{2}},
\end{equation}
with $\tilde{A}=5\sqrt{3}-4\sqrt{5}$ and $\tilde{B}=4\sqrt{3}-3\sqrt{5}$.

As in the van der Waals's gas, here $C_{v}=0$, this makes the line element (\ref{ruppe}) degenerated, and through $\tilde{\mathcal{R}}$ one can see that this value of the heat capacity at constant volume, constitutes a curvature singularity. A way to solve this issue is to introduce the so-called normalized scalar curvature \cite{Wei:2015iwa}
\begin{equation}
\tilde{\mathcal{R}}_{N}=C_{v}\tilde{\mathcal{R}}.
\end{equation}
In this form (\ref{curvature}) becomes
\begin{equation}\label{curvature1}
    \tilde{\mathcal{R}}_{N}(\tilde{v},\tilde{T})=\frac{\left(\sqrt{3}-\tilde{A}\tilde{v}^{4}\right)\left(\sqrt{3}-\tilde{A}\tilde{v}^{4}-2\sqrt{5}\tilde{v}\tilde{T}+\tilde{B}\tilde{v}^{5}\tilde{T}\right)}{2\left(\sqrt{3}-\tilde{A}\tilde{v}^{4}-\sqrt{5}\tilde{v}\tilde{T}+\tilde{B}\tilde{v}^{5}\tilde{T}\right)^{2}},
\end{equation}
eliminating the singularity at $C_{v}=0$. However, there is a double point at 
\begin{equation}
    \tilde{T}_{\text{div}}=\frac{12+3\sqrt{15}+\sqrt{15}\tilde{v}^{4}}{15\tilde{v}+4\sqrt{15}\tilde{v}-3\tilde{v}^{5}}
\end{equation}
where $\tilde{\mathcal{R}}_{N}$ diverges. What is more, at the critical point, its behavior is 
\begin{equation}
    \lim_{(\tilde{v},\tilde{T})\rightarrow (1,1)} \tilde{\mathcal{R}}_{N}(\tilde{v},\tilde{T})\rightarrow -\infty.
\end{equation}
In the middle panel of Fig. \ref{fig3} it is shown the trend of the reduced normalized curvature versus the reduced volume. For small reduced volume, this scalar is positive in nature, which means that the system is under a repulsive interaction. Then, as the reduced volume increases, the reduced normalized curvature becomes negative in nature, implying an attractive interaction. Finally, this scalar again takes positive values for large enough reduced volumes, leading to a repulsive interaction among the particle fluids. Interestingly, there are two divergent points where the scalar curvature goes to negative infinity. With
an increase in temperature, these two divergent points get close and coincide  at $\tilde{T}=1$ (see black line in the middle panel of Fig. \ref{fig3}). This behavior of the reduced normalized curvatures accounts for the thermodynamics phase transition.

Regarding the reduced temperature (coexistence locus), the reduced normalized curvature is exhibited in the right panel of Fig. \ref{fig3}. Here, the large phase (blue line) shows (from right to left) regions where the reduced normalized curvatures go from repulsive, attractive, and repulsive zones. On the other hand, the small phase (red line) is negative (attractive interaction). There is a divergent behavior at $\tilde{T}=1$, and finally, for $\tilde{T}<1$, it becomes positive again (repulsive interaction). 
So, the attractive behavior and the divergent points are in the coexistence phase. These regions are excluded because the equation of
state is invalid in the coexistence phase. So, there is only an attractive interaction among the fluid molecules.

Now, it is important to examine the critical exponent of $\tilde{\mathcal{R}}_{N}$ near the critical point along the coexistence curves for small phase (sp) and large phase (lp), which can give us some universal properties of the system. In general, it satisfies
\begin{equation}
    \tilde{\mathcal{R}}_{N}\sim -(\tilde{T}-1)^{\hat{\epsilon}}. 
\end{equation}
Evaluating (\ref{curvature1}) at (\ref{smalllarge}) and keeping the leading order, one gets
\begin{equation}
    \tilde{\mathcal{R}}_{N}(\text{sp})= \tilde{\mathcal{R}}_{N}(\text{lp})= -\frac{1}{8}(\tilde{T}-1)^{-2}+\mathcal{O}(\tilde{T}-1)^{-3/2}.
\end{equation}
Therefore, from the above expansion, as $\tilde{T}\rightarrow 1$ 
\begin{equation}
     \tilde{\mathcal{R}}_{N}(\tilde{T}-1)^{2}=-\frac{1}{8}.
\end{equation}
This result confirms that near the critical point $ \tilde{\mathcal{R}}(\tilde{T}-1)^{2}C_{v}$ has a constant value $-1/8$ and critical exponent $\hat{\epsilon}=2$. Again, these values agree to those found for a van der Waals's gas \cite{Wei:2015iwa}. Nevertheless, this system drifts from the usual van der Waals gas behavior, because it comes from a relativistic entropy.

\section{Conclusions}\label{sec4}

This paper studies modified Friedmann equations using Kaniadakis's entropy, thermodynamics phase transitions, and the microstructure of the FLRW cosmological model. 

As a first approach to the problem, we have worked at leading order on the parameter $K$ for the EoS (\ref{eos2}). This allows for obtaining analytical solutions for the system.
(\ref{criti}). The results show that this model is driven by a stiff matter distribution, subject to a negative absolute temperature (\ref{criticalpoints}). These facts reveal that cosmology corresponds to an expanding cosmological era where the AH is an outer-past surface. 

Concerning the thermodynamics behavior of this toy model, here the system undergoes a first-order phase transition. This phenomenon occurs for values smaller than the critical temperature $T<T_{c}$. In contrast, for those values above the critical one ($T>T_{c}$), the system does not present any drastic change (see left and middle panels in Fig. \ref{fig1} and Fig. \ref{fig2} where Gibbs's free energy displays the swallowtail shape). This behavior is novel because, in comparison with van der Waals's fluid, it is quite similar, but is it coming from relativistic corrections.

The behavior near the critical point shows that the critical exponents acquire the same values as van der Waals's fluid in the mean-field theory. Also, there are small and large phases, as shown in the left panel of Fig \ref{fig3}. From the point of view of Ruppeiner's geometry, the normalized scalar curvature also has the corresponding van der Waals's critical exponent. Besides expanding this object around the small and large phases obtains the universal constant $-1/8$ at the leading order. At the coexistent phase, the normalized scalar curvature is positive. It becomes negative as $\tilde{v}$ approaches the critical value. Then, it is positive again for large $\tilde{v}$. From previous studies, we know that $\tilde{\mathcal{R}}_{N}>0$ means repulsive interaction and $\tilde{\mathcal{R}}_{N}<0$ attractive interaction. Here, the coexistence phase presents both types of interaction among the particles of the system (see middle panel of Fig. \ref{fig3}). However, these states are invalid or covered by the thermodynamics description of the EoS (\ref{reduceseos}). In terms of the reduced temperature at the small and large phases, the trend of $\tilde{\mathcal{R}}_{N}$ is depicted in the right panel of \ref{fig3}. As can be appreciated, there are two divergent points for the large phase, while there is only one divergent point around the critical point for the small one. Furthermore, the normalized scalar curvature changes in sign only once for the small phase, coming from attractive to repulsive interaction.

To conclude this work, it is worth mentioning that Kaniadakis's entropy introduces interesting and intriguing modifications to the Friedmann cosmological model from a pure standard thermodynamics description.  It is important to highlight that the expansion given by Eq. (\ref{eos2}) is valid in all its domains, that is, for those epochs where $K$ contributions prevail (early Universe) and also for the current era where $K\rightarrow 0$. However, to further validate the latter case, one needs to guarantee that $K$ decreases fast enough to overcome the increasing behavior of $S_{\text{AH}}$ and, in this way, drop out higher-order terms. On the other hand, for the early Universe, both $K$ and $S_{\text{AH}}$ are small, then $KS_{\text{AH}}$ is small, allowing again to keep only the leading order in the expansion (\ref{eos2}) to explore the thermodynamics behavior of the system analytically.
Therefore, applying Kaniadakis's cosmology to the Universe's thermal history is still an open issue. We also need to consider the possibility of a running Kaniadakis's parameter (energy-dependent) to trace all the Universe's stages. Nevertheless, further studies are necessary to corroborate this. This issue shall be addressed elsewhere.

\section*{ACKNOWLEDGEMENTS}
The authors thank the anonymous referees for their useful
remarks, which helped improve the present work's clarity and quality. J. Housset, J. Saavedra and F. Tello-Ortiz acknowledge to grant FONDECYT N°1220065, Chile.
F. Tello-Ortiz acknowledges VRIEA-PUCV
for financial support through Proyecto Postdoctorado 2023 VRIEA-PUCV.

\bibliography{biblio.bib}
\bibliographystyle{elsarticle-num}

\end{document}